# Second harmonic generation from metallo-dielectric multilayer photonic band gap structures

M.C. Larciprete[1], A. Belardini[1], M.G. Cappeddu[2,3], D. de Ceglia[3,4], M. Centini[1], E. Fazio[1], C. Sibilia[1], M.J. Bloemer[3], M. Scalora[3]

1) INFM at Dipartimento di Energetica, Università di Roma "La Sapienza" Via A. Scarpa 16, 00161, Rome, Italy

2) Dipartimeno dei Materiali, Università di Roma, via Eudossiana 18, I-00184 Rome, Italy

3) Charles M. Bowden Research Facility, US Army RDECOM, AMSRD-AMR-WS-ST, Redstone Arsenal, AL 35898

4) Dipartimento di Elettrotecnica ed Elettronica, Politecnico di Bari, Via Orabona 4, 70124 Bari, Italy

## Abstract

We experimentally and theoretically investigate the second order nonlinear optical response of metallo-dielectric multilayer structures composed of Ag and $Ta_2O_5$ layers, deposited by magnetron sputtering. Second harmonic generation measurements were performed in reflection mode as a function of incidence angle, using femtosecond pulses originating from a Ti:Sapphire laser system tuned at λ=800 nm. The dependence of the generated signal was investigated as a function of pump intensity and polarization state. Our experimental results show that the conversion efficiency from a periodic metallo-dielectric sample may be enhanced by at least a factor of 30 with respect to the conversion efficiency from a single metal layer, thanks in part to the increased number of active surfaces, pump field localization and penetration inside the metal layers. The conversion efficiency maximum shifts from 70° for the single silver layer down to approximately 55° for the stack. The experimental results are found to be in good agreement with calculations based on coupled Maxwell-Drude oscillators under the action of a nonlinear Lorentz force term.

**Introduction**

The study of second order nonlinear optical effects in centrosymmetric media has intrigued researchers since the early days of nonlinear optics because it displays peculiar dynamical characteristics with respect to more conventional, non-centrosymmetric media. These peculiarities arise because the electric dipole term vanishes when inversion symmetry is present in the lattice structure. The situation just described applies to most metals, since they posses simple cubic crystal structure. The linear optical susceptibility in metals thus typically includes contributions from conduction [1] and bound [2] electrons.

In 1964, using a classical oscillator electron model, Adler pointed out [3] that in centrosymmetric media the SH source terms consist of a magnetic dipole term, originating from the Lorentz force on the electrons, and of an electric quadrupole contribution, through the Coulomb force. Subsequently, Jha [4] used a free electron gas model to show that the quadrupole source term was equivalent to a nonlinear surface contribution. He was the first to propose that SHG in metals could be explained by two phenomenological contributions having the form $\mathbf{P}_{2\omega} = \alpha \mathbf{E}_\omega \nabla \cdot \mathbf{E}_\omega + \beta \mathbf{E}_\omega \times \mathbf{H}_\omega$, where α and β are predetermined, frequency-dependent coefficients [4] that multiply a surface and a volume (or bulk) contribution, respectively.

The first experimental results outlined by Brown and coworkers appeared to confirm the existence of two SH source terms by two-fold excitation of a silver layer by a pump linearly polarized normal and parallel to the plane of incidence [5,6], which can in turn excite volume or surface sources, respectively. Later, Bloembergen and Shen [7] noted that the SHG reported in references [5,6] was most likely due to contributions from core electrons, as agreement with theory is obtained only when both free and bound electron contributions are considered [8].

Meanwhile, more experimental progress was made as additional metals and configurations were explored. For example, SHG was reported in total internal reflection from a film immersed in a denser medium [9], from opaque films deposited on glass prisms [10], and from thin metal films sandwiched between two dielectric layers [11], where coupling with surface plasmons and enhanced SHG was also observed. Several phenomenological approaches were proposed in order to fit the experimental data, as exemplified by the work of Rudnick and Stern [12], who also used two parameters to describe the nonlinear SH source currents.

Recently, the use of metals for applications in the optical range has grown together with the interest in their applications in nonlinear optics. It has been shown that thin metal films can be included into multilayer structures to achieve high transmittance in the visible range and beyond, despite the high imaginary part of the index of refraction typical of metals [13, 14]. These metallo-dielectric, multilayer structures, also known as transparent metals, consist of both periodic and symmetric structures, composed by the alternation of metallic and dielectric or semiconductor layers. Ordinarily, light can propagate inside a thick metal layer only up to a small distance (this distance is known as the skin depth, which for typical metals ranges between 5 and 10nm in the visible range) beyond which it is mostly attenuated. In the case of transparent metals, the skin depth limit is overcome. A resonant tunneling mechanism renders hundreds of nanometers of metal transparent, and allows both TE- and TM-polarized fields to become localized inside both the metal and dielectric layers, without the usual detriments of absorption associated with the high imaginary index component. These structures thus turn out to be an extraordinary instrument to access and enhance the nonlinear optical response of nonlinear layers [15], and in particular, the second [16] and third [17, 18] order optical nonlinearities of metals. This latter feature is particularly interesting when investigating second order nonlinear effects because most metals present centrosymmetric crystal structure, so that the SH source terms arise from magnetic dipole and electric quadrupole contributions [19]. In the case of bulk metals, surface effects play a dominant role and are responsible for most of the generated signal. On the other hand, in non-centrosymmetric crystals surface effects may become significant and contribute to SHG only when the films are either amorphous or very thin. Thus the opportunity of including several metal layers into stacks where the light can become strongly localized opens new vistas, and may broaden the range of likely applications due to the possibility of increasing the number of active surfaces and volume contributions for the enhancement of SHG.

**Sample Preparation**

In our experiments, we measured the second harmonic signal in the blue spectral region (400 nm) generated by different $Ag/Ta_2O_5$ multilayer structures. The pump consisted of pulses approximately 150fs in duration, originating from a Titanium:Sapphire pulsed laser system centered at a wavelength of 800nm, and having a repetition rate of ~1 kHz. The samples were grown on glass substrates by means of a magnetron sputtering system [10]. Magnetron sputtering is a well-established technique for thin film deposition that allows the deposition of several materials without breaking vacuum, and thus it is well-suited for the fabrication of multilayer structures. In what follows, we will describe the details of sample preparation and realization, the experimental setup used to conduct the SHG measurements, and the theoretical model that we adopted for the analysis of the experimental data.

The sputtering chamber was evacuated by a turbo molecular pump to a final pressure of approximately $10^{-7}$ bar. Films were deposited onto 1 mm thick, optically flat ($\lambda/20$) glass substrates that had been previously cleaned with ethylic alcohol to remove dust and other possible organic contaminants, and subsequently by exposing them to dry air flux. Substrates were placed on a sample holder that rotated at 20 revolutions per minute, in order to increase film uniformity. The electrical power applied to the electrodes was set to 150W for $Ta_2O_5$, and 100W for Ag. Under these deposition conditions, the growth rate was found to be 2.5 Å/s for $Ta_2O_5$, and 9.5 Å/s for Ag layers. During deposition, the chamber temperature was monitored through a thermocouple placed behind the substrate, and it was found that it rose by ~10°C from its initial value of ~27°C. After deposition was completed, a linear optical characterization of the samples was carried out. The transmittance spectra were recorded at normal incidence in the visible-NIR range by spectro-photometric technique (Lambda 19 Spectrophotometer by Perkin-Elmer) [21]. In Fig.(1), we report the transmission spectrum obtained from a sample consisting of 5 periods of Ag(20nm)/$Ta_2O_5$(124nm). The experimental curves were reconstructed using a standard transfer-matrix algorithm. The optical constants for Ag used to fit the data were taken from Palik [22], and the optical constants for $Ta_2O_5$ were fitted from experimental reflectance data of a single $Ta_2O_5$ film deposited on Si with a Filmetrics reflectometer having a lower wavelength range of 600 nm. The $Ta_2O_5$ optical constants were extrapolated for the shorter wavelengths. The linear transmittance function is reproduced with reasonably good agreement, and it is plotted in Fig.(1). The low transmittance at 400 nm for the experimental data is likely a result of the absorption in the $Ta_2O_5$ that increases for shorter wavelengths.

**Second Harmonic Generation**

The second harmonic signal was measured to evaluate the conversion efficiency [23] of the multilayer samples, and compared to the conversion efficiency of a single metal layer. The technique consists in measuring the reflected SH signal for a given polarization state of both the fundamental input and SH output beams. The schematic representation of the experimental setup used for the measurements is shown in Figure 2. The main beam was focused onto the sample by a lens having a 150-mm focal length. The polarization states of both the fundamental and the generated beams may be set via a half-wavelength plate placed before the lens, while a polarizer for signal analyzing is placed before the detector. In order to remove the SH signal produced by the plate's crystals, due to the short pulse duration, a long pass filter (GG495, Thorlabs) was placed after the half-wave plate. The sample was placed on a rotational stage which allowed setting of the incidence angle with a resolution of 0.5 degrees. The transverse profile of the fundamental beam was measured and resulted to be Gaussian with a waist ranging from 400 to 700 μm, depending on the sample-to-focus distance. This distance could be varied by adjusting the lens's position, and by repositioning the rotational sample holder in the center of the beam.

After being reflected by the sample, the fundamental and second harmonic beams were sent through a glass prism and thus separated. A set of dichroic filters was then used to further suppress any residual and scattered FF, thus ensuring that only the SH beam was directed to the photomultiplier tube, and then analyzed by a 500 MHz digital oscilloscope. The photomultiplier output was then fed into a box-car averager, increasing the signal-to-noise ratio. The calibration curve of the photomultiplier response was accurately performed with a reference BBO crystal. When necessary, detector saturation was prevented by using a set of linear neutral density filters whose transmittance value was taken into account in the data processing. The incident FF light was strictly plane polarized, and the polarization state introduced by the half-wave plate was checked by preliminary calibration carried out with a second crossed polarizer used to analyze the polarization of FF beam before the sample, in order to avoid undesired components of the FF electric field. Experimental measurements show that the largest signal is recorded when the polarization of the fundamental beam is set to $\hat{p}$, while the SH signal is always $\hat{p}$-polarized.

The first set of measurements was done by increasing the FF peak power, in order to check for a quadratic dependence of the SH signal on the FF peak power. We investigated a number of periodic and symmetric samples (the latter having dielectric entry and exit layers), and in all cases we found that the generated power at 400 nm has a quadratic dependence on the FF squared peak power. The next step was to investigate the laser light polarization dependence of the SH signal by varying the angle ϕ between FF polarization state and the plane of incidence. According to the arguments presented in e.g. reference [6], when the FF electric field component in the plane of incidence is zero (s-polarized) there should be only a bulk nonlinear contribution excited through the Lorentz force. This contribution, which is directed longitudinally, in the same direction as the wave vector (radiation pressure), can still propagate in the presence of a boundary, i.e. for nonzero incidence angle. On the other hand, when the FF is polarized in the plane of incidence, the SH contribution is predominantly of surface origin, arising from the induced nonlinear currents, or equivalently from longitudinal field discontinuities.

In order to measure the SH dependence on the FF polarization direction, the polarization of the FF was varied between 90° and 0°, at a fixed incidence angle of 45°. By analyzing the SH polarization direction we found that the SH light is always polarized in the plane of incidence, i.e. $\hat{p}$-polarized. In Fig.(3) we report the curves of the SH signal as a function of the polarization direction of FF beam, ϕ. The figure shows that the SH signal does not go to zero when the FF is polarized normally with respect to the plane of incidence, thus indicating that in the multilayer structure the nonlinear process is excited also via the bulk term of the nonlinearity. Thus, by introducing the parameter M as the ratio of the relative second harmonic signal measured for ϕ=90° and ϕ=0° respectively, $M = Signal_{90°}/Signal_{0°}$, our experimental curves reveal a value of M~0.15. This value should be contrasted with the values previously reported in the seminal work of reference [6] for a single Ag layer, which was found to be in the range 0.02 to 0.06. This result means that SHG from volume contributions is not negligible, and that volume sources can be excited in multilayer stacks by choosing suitable dielectric layer thicknesses between two consecutive metal layers to form a transparent metallo-dielectric photonic band gap structure.

**Theoretical Model**

Before discussing the angular dependence of second harmonic generation that we measured we describe the theoretical model that we adopted to predict second harmonic generation from centrosymmetric materials. Although Sipe's hydrodynamic model [24] is widely used to analyze experimental data [10, 11, 25], we assume that the metal consists of a free electron gas described by the Drude model, under the action of a driving electromagnetic field [1, 26]. Under these conditions, longitudinal and transverse nonlinear currents arise under the action of the nonlinear Lorentz force [26]. It is widely known that metallic data cannot generally be fitted throughout the visible and near IR ranges by a single set of $(\gamma, \omega_p)$ parameters, which stand for damping coefficient and plasma frequency, respectively. Actual metal data, as exemplified in Palik's handbook [22], displays core electron contributions well into the visible range, so that a more complex system of equations must be used. One possible way to proceed is to supplement the simple Drude model with one or more Lorentz oscillator equations that describe core electrons. Since for the moment we are interested in two frequencies only, FF and SH, a simpler way forward consists of fitting the data using the Drude model and two different sets of $(\gamma, \omega_p)$ parameters, each set fitted to the frequency of interest. In doing so we also seek to match the slope of the complex dielectric function at each frequency, in order to impart the correct group velocities to both the FF and SH frequencies. In principle, this procedure can be repeated for an arbitrary number of harmonics, but it becomes more difficult to simultaneously fit both the dielectric function and its derivative in proximity of the plasma frequency. We note however that for structures only a fraction of a wavelength thick fitting the group velocity is not as important as fitting the dielectric function, since propagation distances are extremely small [18].

In Gaussian units, the system of equations we aim to solve is thus as follows:

$$\nabla \times \mathbf{E} = -\frac{1}{c}\frac{\partial \mathbf{B}}{\partial t} \qquad \nabla \times \mathbf{H} = \frac{1}{c}\frac{\partial \mathbf{E}}{\partial t} + \frac{4\pi}{c}\frac{\partial \mathbf{P}}{\partial t}$$
$$\ddot{\mathbf{P}} + \gamma \dot{\mathbf{P}} = \frac{\omega_p^2}{4\pi}\mathbf{E} + \frac{e}{mc}\dot{\mathbf{P}} \times \mathbf{H} \qquad (1)$$

We assume a right-handed coordinate system, and $\hat{p}$-polarized (TM) pump and second harmonic fields of the type:

$$\mathbf{E} = \begin{pmatrix} \mathbf{j}\left(E_y^\omega e^{-i\omega t} + \left(E_y^\omega\right)^* e^{i\omega t} + E_y^{2\omega} e^{-2i\omega t} + \left(E_y^{2\omega}\right)^* e^{2i\omega t}\right) + \\ \mathbf{k}\left(E_z^\omega e^{-i\omega t} + \left(E_z^\omega\right)^* e^{i\omega t} + E_z^{2\omega} e^{-2i\omega t} + \left(E_z^{2\omega}\right)^* e^{2i\omega t}\right) \end{pmatrix}$$

$$\mathbf{H} = \mathbf{i}\left(H_x^\omega e^{-i\omega t} + \left(H_x^\omega\right)^* e^{i\omega t} + H_x^{2\omega} e^{-2i\omega t} + \left(H_x^{2\omega}\right)^* e^{2i\omega t}\right) \qquad (2)$$

The corresponding macroscopic polarization is given by:

$$\mathbf{P} = (P_y \mathbf{j} + P_z \mathbf{k}) = \begin{pmatrix} \mathbf{j}\left(P_y^\omega e^{-i\omega t} + \left(P_y^\omega\right)^* e^{i\omega t} + (P_y^{2\omega} e^{-2i\omega t} + \left(P_y^{2\omega}\right)^* e^{2i\omega t}\right) + \\ \mathbf{k}\left(P_z^\omega e^{-i\omega t} + \left(P_z^\omega\right)^* e^{i\omega t} + P_z^{2\omega} e^{-2i\omega t} + \left(P_z^{2\omega}\right)^* e^{2i\omega t}\right) \end{pmatrix} \qquad (3).$$

The envelope functions $E_y^{\omega,2\omega}$, $E_z^{\omega,2\omega}$, $H_x^{\omega,2\omega}$, $P_y^{\omega,2\omega}$, $P_z^{\omega,2\omega}$ contain implicit spatial dependences that for simplicity have been omitted. In addition, the envelope functions are not assumed to be slowly varying, as no approximations are made when the fields and polarizations are substituted into Maxwell's equations. For second harmonic generation, substitution of Eqs.(2-3) into Eqs.(1) results in a system of fourteen coupled differential equations for the fields, polarizations, and corresponding currents, which in scaled form are written as follows for the pump:

$$\frac{\partial H_x^\omega}{\partial \tau} = i\beta\left(H_x^\omega + E_z^\omega \sin\theta_i + E_y^\omega \cos\theta_i\right) - \frac{\partial E_z^\omega}{\partial \tilde{y}} + \frac{\partial E_y^\omega}{\partial \xi}$$

$$\frac{\partial E_y^\omega}{\partial \tau} = i\beta\left(E_y^\omega + H_x^\omega \cos\theta_i\right) + \frac{\partial H_x^\omega}{\partial \xi} - 4\pi(J_y^\omega - i\beta P_y^\omega)$$

$$\frac{\partial E_z^\omega}{\partial \tau} = i\beta\left(E_z^\omega + H_x^\omega \sin\theta_i\right) - \frac{\partial H_x^\omega}{\partial \tilde{y}} - 4\pi(J_z^\omega - i\beta P_z^\omega)$$

$$\frac{\partial J_y^\omega}{\partial \tau} = (2i\beta - \gamma_\omega)J_y^\omega + (\beta^2 + i\gamma_\omega \beta)P_y^\omega + \frac{\pi\omega_{p,\omega}^2}{\omega_r^2}E_y^\omega$$

$$+ \frac{e\lambda_0}{mc^2}\left[\left((J_z^\omega)^* + i\beta(P_z^\omega)^*\right)H_x^{2\omega} + (J_z^{2\omega} - 2i\beta P_z^{2\omega})(H_x^\omega)^*\right]$$

$$\frac{\partial J_z^\omega}{\partial \tau} = (2i\beta - \gamma_\omega)J_z^\omega + (\beta^2 + i\gamma_\omega \beta)P_z^\omega + \frac{\pi\omega_{p,\omega}^2}{\omega_r^2}E_z^\omega$$

$$- \frac{e\lambda_0}{mc^2}\left[\left((J_y^\omega)^* + i\beta(P_y^\omega)^*\right)H_x^{2\omega} + (J_y^{2\omega} - 2i\beta P_y^{2\omega})(H_x^\omega)^*\right]$$

$$J_y^\omega = \frac{\partial P_y^\omega}{\partial \tau} \qquad J_z^\omega = \frac{\partial P_z^\omega}{\partial \tau} \qquad , \quad (4)$$

and as follows for the SH:

$$\frac{\partial H_x^{2\omega}}{\partial \tau} = 2i\beta\left(H_x^{2\omega} + E_z^{2\omega}\sin\theta_i + E_y^{2\omega}\cos\theta_i\right) - \frac{\partial E_z^{2\omega}}{\partial \tilde{y}} + \frac{\partial E_y^{2\omega}}{\partial \xi}$$

$$\frac{\partial E_y^{2\omega}}{\partial \tau} = 2i\beta\left(E_y^{2\omega} + H_x^{2\omega}\cos\theta_i\right) + \frac{\partial H_x^{2\omega}}{\partial \xi} - 4\pi(J_y^{2\omega} - 2i\beta P_y^{2\omega})$$

$$\frac{\partial E_z^{2\omega}}{\partial \tau} = 2i\beta\left(E_z^{2\omega} + H_x^{2\omega}\sin\theta_i\right) - \frac{\partial H_x^{2\omega}}{\partial \tilde{y}} - 4\pi(J_z^{2\omega} - 2i\beta P_z^{2\omega})$$

$$\frac{\partial J_y^{2\omega}}{\partial \tau} = (4i\beta - \gamma_{2\omega})J_y^{2\omega} + (4\beta^2 + i\gamma_{2\omega}2\beta)P_y^{2\omega} + \frac{\pi\omega_{p,2\omega}^2}{\omega_r^2}E_y^{2\omega} + \frac{e\lambda_0}{mc^2}(J_z^\omega - i\beta P_z^\omega)H_x^\omega$$

$$\frac{\partial J_z^{2\omega}}{\partial \tau} = (4i\beta - \gamma_{2\omega})J_z^{2\omega} + (4\beta^2 + i\gamma_{2\omega}2\beta)P_z^{2\omega} + \frac{\pi\omega_{p,2\omega}^2}{\omega_0^2}E_z^{2\omega} - \frac{e\lambda_0}{mc^2}(J_y^\omega - i\beta P_y^\omega)H_x^\omega$$

$$J_y^{2\omega} = \frac{\partial P_y^{2\omega}}{\partial \tau} \qquad J_z^{2\omega} = \frac{\partial P_z^{2\omega}}{\partial \tau}$$

. (5)

We have chosen $\lambda_r = 1\mu m$ as the reference wavelength, and have adopted the following scaling: $\xi = z/\lambda_r$ and $\tilde{y} = y/\lambda_r$ are the scaled longitudinal and transverse coordinates, respectively; $\tau = ct/\lambda_r$ is the time in units of the optical cycle; $\beta = 2\pi\tilde{\omega}$ is the scaled wave vector; $\tilde{\omega} = \omega/\omega_r$ is the scaled frequency, and $\omega_r = 2\pi c/\lambda_r$, where $c$ is the speed of light in vacuum. $\theta_i$ is the angle of incidence of the pump with respect to the normal direction. The magnitude of the coupling coefficient in the Lorentz force term is evaluated in Gaussian units: $\frac{e\lambda_0}{mc^2} = \frac{(-4.8\times10^{-10})\times(10^{-4})}{(9.1\times10^{-28})\times(9\times10^{20})} = -5.9259\times10^{-8}(cgs)$. It is known that the effective electron mass in silver is close to the bare electron mass [27]. In the context of Eqs.(4-5) above and so for simplicity, we choose $m$ to the bare electron mass. The linear dielectric response of silver is assumed to be Drude-like, as follows: $\varepsilon(\tilde{\omega}) = 1 - \frac{\omega_P^2}{\tilde{\omega}^2 + i\gamma\tilde{\omega}}$. At 800nm, the data [22] is fitted using the set of parameters: $(\gamma_\omega, \omega_{p,\omega})=(0.06, 6.73)$, and at 400nm we have: $(\gamma_{2\omega}, \omega_{p,2\omega})=(0.33, 5.51)$. The incident magnetic field was assumed to be Gaussian of the form: $\mathcal{H}_x(\tilde{y}, \xi, \tau=0) = H_0 e^{-[(\xi-\xi_0)^2 + \tilde{y}^2]/w^2}$, with similar expressions for the transverse and longitudinal electric fields. Finally, we fitted our measured $Ta_2O_5$ data using a standard Taylor expansion (see caption of Fig.(1) for the $Ta_2O_5$ data used in the calculations), and inserted the relevant parameters in the model. As can be readily ascertained from the equations of motion, leaving aside the absence of third and higher harmonics, the model quite completely and exhaustively describes the interaction of an incident pump pulse with a generic centrosymmetric material, including a multilayer stack. The set of coupled equations (4) and (5) are integrated using the fast Fourier transform-based pulse propagation technique to propagate the fields [18], and a simple, second-order accurate predictor-corrector algorithm to advance the temporal solutions of the currents and polarizations.

**Results and Discussion**

In Fig.(4) we depict typical incident and scattered pump and SH pulses. Details about grid size and other discretization parameters are found in the caption. In Fig.(5) we report our predictions of SH conversion efficiency η vs. incident angle using our model, for a single 20nm-thick silver layer. We define conversion efficiency as the ratio of either transmitted or reflected SH energy divided by incident pump energy. The results are consistent with the results found throughout the literature, that is, maximum conversion efficiency for Ag occurs at approximately 70° on reflection. Our results suggest that the transmitted SH signal is also peaked at 70°. For a peak pump intensity of approximately 6 GW/cm$^2$, the predicted conversion efficiency upon reflection is $\sim 1.4 \times 10^{-11}$. The calculated conversion efficiencies quickly converge for pulses only a few tens of femtoseconds in duration and having a relatively small spot size because, unlike the multilayer stack, the single metal layer presents no significant structure in the transmission function.

In Fig.(6) we show the predicted transmitted and reflected SHG efficiencies η as a function of incident angle, for the 5-period, metallo-dielectric stack depicted in Fig.(1). The incident field is Gaussian in space and time, with a spot size approximately 30μm wide and 150fs duration (~1/e width). The input peak intensity is taken to be roughly 6 GW/cm$^2$. A maximum conversion efficiency of $\sim 2.8 \times 10^{-10}$ (an improvement of a factor of ~20 compared to the single metal layer) is thus predicted for the multilayer stack, and it occurs at ~55° for both transmission and reflection coefficients. The essential results suggest that SHG is most efficient for pulses that are long enough to resolve the features of the transmission resonances shown in Fig.(1), when pump penetration depth inside the metal and absorption are maximized, and when the longitudinal component of the electric field displays the largest discontinuities. In Fig.(7) we show a plot of the corresponding forward and backward SH conversion efficiency as a function of pulse duration. Longer, narrower bandwidth pulses tend to better localize inside the stack, leading to higher local field intensities. These findings are consistent with the results discussed in reference [16], where an illustrative, simplified model of SHG from a metallo-dielectric stack was discussed in the context of normal incidence and uniformly distributed nonlinear dipoles.

Field localization, the bandwidth of the incident pulse, and phase matching conditions are usually of central importance in the study of SHG in structures of finite length [28, 29]. It has been demonstrated that it is possible for either phase matching conditions [29] or field localization effects [28] to dominate the conversion process, depending on structure size and field overlap. In typical symmetric and asymmetric transparent metallo-dielectric stacks that are less than one wavelength thick, the fields become localized inside both the metal and the dielectric layers [16, 17, 30]. The calculations consistently show that linear pump and nonlinear SH currents and dipoles are present at each metal surface with relatively high field intensities inside each layer and longitudinal field discontinuities at every interface. We illustrate this in Fig.(8), where we show transverse and longitudinal electric field components inside the stack, for 55º angle of incidence, when the peak of the incident pulse reaches the stack. The figure shows that the presence of multiple active surfaces and the light's ability to penetrate and dwell inside the metal help the enhancement of SHG. It is also evident that the transverse component of the electric field is also relatively intense inside each metal layer thanks to the resonance tunneling phenomenon. As a result, the contribution of volume sources becomes an integral part of the interaction. Nevertheless, some inherent uncertainties about the relative importance of volume and surface contributions remain, due to the mere complexity of the theoretical model, i.e. Eqs.(4-5) above, making it difficult, at least for now, to determine the relative importance of volume and surface effects. A few simple examples should suffice to highlight the complexity of the problem.

Our results suggests that a number of factors combine to yield enhanced SHG, namely: (i) field localization inside the metal layers; (ii) pulse duration, which is intimately connected to the first point; (iii) tuning at frequencies near the long wavelength band edge, where field penetration depth inside the metal and linear absorption are maximized; and (iv) the ability to establish field discontinuities and nonlinear dipole distributions throughout the stack. All these factors may be termed as volume contributions that have no counterpart in isolated, relatively thick metal layers. It is relatively easy to establish that in isolated metal layers surface effects are directly responsible for most of the observed SHG. One may show this by calculating the field profiles, and by monitoring the difference between the intensities just inside and just outside the entry surface. In Fig.(9) we plot such a field discontinuity as a function of incident angle for an incident field of unitary amplitude. An examination of the figure and only a cursory comparison with Fig.(5) suffices to confirm a direct correlation between surface effects and large SH conversion efficiencies, as both display the same angular dependence. Thinner metal layers display a similar response.

We now examine a simple example that illustrates how volume effects may indeed dominate over surface effects in the case of transparent metal stacks. We examine the field profiles for the same periodic structure we have considered above, except that now we turn the structure around so that the field is incident on the metal instead of the dielectric layer. Of course, the linear transmittance properties of the stack do not change, regardless of the direction of approach. However, if light is incident on the metal layer, a large field discontinuity is recorded at the metal interface rather than the dielectric interface. All things being equal, the large field discontinuity present in one stack does not at all improve SHG conversion efficiencies. In fact, reversing the stack yields slightly lower reflection SH efficiency, with even smaller *transmitted* SHG. As a result, one might surmise that volume contributions must be compensating the evident surface effect that characterizes the sample if it is positioned in a way that light is incident on the metal side. Similar results were obtained for a variety of stacks.

These results seem to suggest that volume contributions may indeed play a role more pronounced than one may be able to presently discern. However, to arrive at such a definitive conclusion, one should construct a model where it is possible to selectively isolate surface from volume contributions, and then integrate the equations of motion to record the effect. Unfortunately, the model we use suggests that inside the transparent metallo-dielectric stacks surface and volume contributions may be inextricably linked, thus making it difficult to distinguish their relative importance as the fields actually penetrate and are relatively intense inside the metal layers.

There are several issues that one must take into consideration when adding, subtracting, thickening, or thinning metal or dielectric layers. For example, adding periods generally increases reflections, shifts and narrows the transmission resonances, and the field becomes better localized inside the dielectric layers because the metal layers act as better mirrors. Thinning the metal layers and increasing their number requires adjustment of the dielectric layer thickness in order to keep the resonance tunneling mechanism operating within a desired wavelength range, and to keep both fields tuned inside a pass band. One might think that having as many metal layers as possible can increase conversion efficiency. This is generally not the case, because volume contributions also come in the form of enhanced linear absorption (as a result of field localization inside the metal), which can overwhelm any nonlinear gain. Therefore, structures that contain many layers actually may perform worse than a single metal layer, as the FF, the SH or both fields may slide into their respective gaps. Finally, it is noteworthy that for relatively large incident angles, such as those we are considering, the scattered SH fields are generated as they propagate sideways along the length of the metal layers, for several tens of microns before they exit the structure, as Fig.4 suggests. This naturally translates into a great deal of effective instantaneous losses, which combine with an instantaneous gain large enough to yield the modest conversion efficiencies that we observe.

These results thus generally suggest that although there is a strong hint that volume contributions may in effect play a role far more important than surface discontinuities, the examples we have investigated, which include periodic and a variety of symmetric, more transmissive stacks, at present suggest that it is difficult to extract their relative importance. When designing the stacks one should be make judicious choices in the selection of the number of layers, their relative thickness, and tune the fields at the long wavelength band edge, at a place of relatively high linear absorption, where the fields are still well-localized inside the metal. At the same time, one should avoid tuning where there is strong feedback, such as at the peak of narrow resonances, which have a tendency to kick the field back into the dielectric layers and reduce nonlinear gain.

One can easily see that the subject is extremely complex and interesting, primarily because it brings us full circle to fundamental questions and issues explored during the early history of nonlinear optics. For this reason alone the subject deserves to be investigated further. Suffice it to say here that the newly acquired ability of the fields to penetrate and dwell inside metal layers combined with the ability to excite multiple metal surfaces changes the dynamical characteristics of SHG in metals, with competing surface and volume contributions. This statement is reflected in our simulations and in our findings, as reported above.

One final point worthy of note should be made about spot size. Although the spot size used in the calculations (~30μm) is significantly smaller compared to that used in the experiments (upward of 500μm), a 30-μm beam width corresponds to a fairly narrow bandwidth of transverse k-vectors that tend to resolve well all features found in the transmission function of Fig.1, for example. In other words, plane wave results are quickly achieved provided the beam waist is taken to be at least several wavelengths wide.

In Fig.(10) we report the measurements performed in reflection mode as a function of the incidence angle for the 5-period $Ta_2O_5$/Ag sample for an input FF intensity of ~6GW/cm$^2$. The polarization of both fundamental and generated beams lies in the plane of incidence. As a comparison, we also plot the measurements obtained for a single Ag layer 20nm thick, obtained under similar experimental conditions. Just as predicted by our model in Figs.(5-6), the signal arising from the multilayer structure displays a maximum value at an incidence angle of ~55°, instead of 70° for the single metal layer. The theoretical predictions for reflections are also plotted in Fig.(10). The experimental data reported in the figure suggests that the SH signal generated inside the metallo-dielectric stack is enhanced by approximately a factor of 30 relative to the maximum conversion efficiency of the single 20nm-thick Ag layer. Given the extreme complexity of the model, as exemplified by Eqs.(4-5), and some uncertainty about the precise peak intensity that reaches the stack, one may objectively state that the agreement between our theory and our experiment is quite good, especially considering that the theoretical model has no adjustable parameters. Other possible sources of uncertainty include small deviations in layer thicknesses, and third order effects inside the metal layers, that may lead to band shifts and nonlinear absorption [17, 18], which the current model does not take into account. Further studies will focus on extending the model to include a third harmonic frequency, third order effects, and the evaluation of conversion efficiency for other geometrical configurations and metals that might further clarify the relative importance and interplay between surface and volume contributions.

**Conclusions**

In summary, we have theoretically and experimentally investigated second harmonic generation from metallo-dielectric, $Ta_2O_5$/Ag multilayer structures, and compared the SH efficiency with that obtained from a single Ag layer. The SH signal was investigated experimentally in reflection mode using femtosecond pulses originating from a Ti:Sapphire laser system tuned at $\lambda$=800 nm. The experimental results are well-explained within the context of a classical model based on coupled Maxwell-Drude oscillators under the action of a nonlinear Lorentz force term arising from a driving electromagnetic field. Measurements performed on different samples show that the generated signal is dependent on the squared intensity of the fundamental beam and is always polarized in the plane of incidence. The theory, which is largely substantiated by our experimental results, suggests that a combination of maximized pump penetration inside the stack, field localization inside the metal, and large longitudinal field discontinuities all contribute to the enhancement of SHG by approximately a factor of 20 relative to a single metal layer, while experimental results suggests an improvement by at least a factor of 30.    It may be possible to achieve larger conversion efficiencies for alternative geometrical configurations and metals.  We believe that while it is clear that isolated metal layers benefit largely from surface effects, the situation is far more complicated in the case of transparent metallo-dielectric stacks, and more studies are needed to eventually unravel the intricacies of the process.  Due to the peculiar transmittance and field localization properties of metallo-dielectric structures, both the fundamental and SH beams can become localized inside the metal layers.  This localization phenomenon thus allows for the possibility of competing surface and volume contributions, a novel effect that was also confirmed by our experimental study.

**Acknowledgements**

We thank Oleg Buganov for experimental support, and Neset Akozbek for useful discussions.

**Figure Captions**

**Figure 1**. (Color online) Measured (○) and calculated (continuous curve) optical transmittance spectra of the periodic multilayer sample, [Ag(20nm)/Ta$_2$O$_5$(124nm)]x5. Our measurements suggest that the dielectric constant of Ta$_2$O$_5$ is linear in the range of interest, and may be approximated as follows: $\varepsilon$(800nm)~4.6+i7x10$^{-5}$, and $\varepsilon$(400nm)~4.75+i10$^{-3}$. For modeling purposes, absorption may be neglected.

**Figure 2.** Experimental setup for second harmonic generation measurements: passive mode locked Ti:Sapphire laser; VA: variable attenuator for varying FH intensity; λ/2 half-wave plate; HP-F: high pass filter; L$_1$ and L$_2$ 150mm focal length lenses; α: incidence angle; P: glass prism; BS: beam stop for removal of the fundamental beam; A: analyzer; PM: photomultiplier tube.

**Figure 3.** (Color online) Second harmonic generated signal as a function of fundamental beam polarization state, measured in reflection mode, at an incidence angle of 45°, for the periodic sample described in Figure 1. ϕ represents the angle between pump beam polarization direction and the plane of incidence, i.e. when ϕ=0° pump beam is $\hat{p}$-polarized while for ϕ=±90° pump beam is $\hat{s}$-polarized. The SH signal was found to be always polarized in the plane of incidence ($\hat{p}$).

**Figure 4.** Typical scattering event where the pump pulse is partially transmitted and reflected from either a single metal layer, or a multilayer stack. The figure shows that both transmitted and reflected SH pulses (upper right) are generally emitted either specularly or in the same direction of approach of the pumping pulse. The simulation of 150fs incident Gaussian pulses is done using a 32x65536 spatial grid. The grid size reflects fine discretization along the longitudinal coordinate, which contains material discontinuties (δξ=0.004λ$_0$), and much coarser discretization along the transverse coordinate (δy=4.096λ$_0$), made possible by the absence of transverse boundaries. The time is discretized in units of $\delta\tau=0.004\lambda/c$. Typical execution times per shot are between 12 and 72 hours, depending on the angle of incidence, on Pentium D 3.4GHz computer.

**Figure 5.** (Color online) Second harmonic transmission (■) and reflection (●) conversion efficiencies η vs. incident angle (left axis) for a single 20nm-thick silver layer. In the simulations, the calculated conversion efficiencies quickly converge for pulses only a few tens of femtoseconds in duration, with a relatively small spot size, while the peak intensity of the incident Gaussian pulse was set to ~6 GW/cm$^2$. Maximum pump absorption may be ascertained by calculating any remaining pump energy (▲, right axis, normalized to unity). The plot shows maximum conversion efficiency does not coincide with maximum pump absorption.

**Figure 6.** (Color online) Predicted transmitted (■) and reflected (●) second harmonic conversion efficiencies η as functions of incident angle, for the 5-period metal-dielectric stack depicted in Fig.(1). Incident pulse duration is ~150 fs, with a spot size approximately 30 microns wide, and peak intensity ~6 GW/cm$^2$. In this case, field localization and pump penetration inside the sample causes nearly 50% of the incident pump energy to be absorbed. Maximum conversion efficiency and maximum absorption angles nearly coincide. On the right axis we show the remaining pump energy (▲, right axis, normalized to unity).

**Figure 7.** (Color online) Conversion Efficiency vs. incident pulse duration. The increase in conversion efficiency is due to the mere improvement of field localization inside the stack, which generally means increased field intensities inside the metal and dielectric layers. On the x-axis, 60 optical cycles corresponds roughly to 150fs.

**Figure 8.** Snapshot of the longitudinal and transverse electric field intensities, inside (longitudinal axis) the multilayer stack, when the peak of the pulse reaches the stack. The pulse is incident from the left at an angle 55°. The dark, thin regions in the background represent the metal layers. The figure shows two essential facts: (i) the transverse electric

field becomes localized inside each metal layers, and is relatively intense; and (ii) the longitudinal electric field displays large discontinuties at each interface, i.e. $|E_z|^2 \to 0$ inside the metal. The combination of these effects leads to substantially increased surface and volume sources, and an overall enhancement of SHG by at least factor of 20 compared to a single Ag layer.

**Figure 9**. (Color online) Magnitude of field discontinuity $\delta|E|^2$ vs. incident angle for a 100nm Ag layer. $\delta|E|^2$ represents the difference between the field intensities just outside and just inside the metal layer. A simple comparison with Fig.(5) reveals that surface effects may be directly correlated to SHG from an isolated metal layer.

**Figure 10.** (Color online) SH efficiency measured in reflection mode from the multilayer $[Ag(20nm)/Ta_2O_5(124nm)]^5$ sample (●), as a function of incidence angle. For comparison, the SH efficiency measured from a single Ag layer 20-nm thick (■) under the same experimental condition is also reported. Theoretical predictions for each experimental curve are also included from Figs.(6) and (7), as specified in the legend. Pump beam intensity was set to ~6 GW/cm$^2$. The FF and SH beams are both polarized in the plane of incidence ($\hat{p}$).

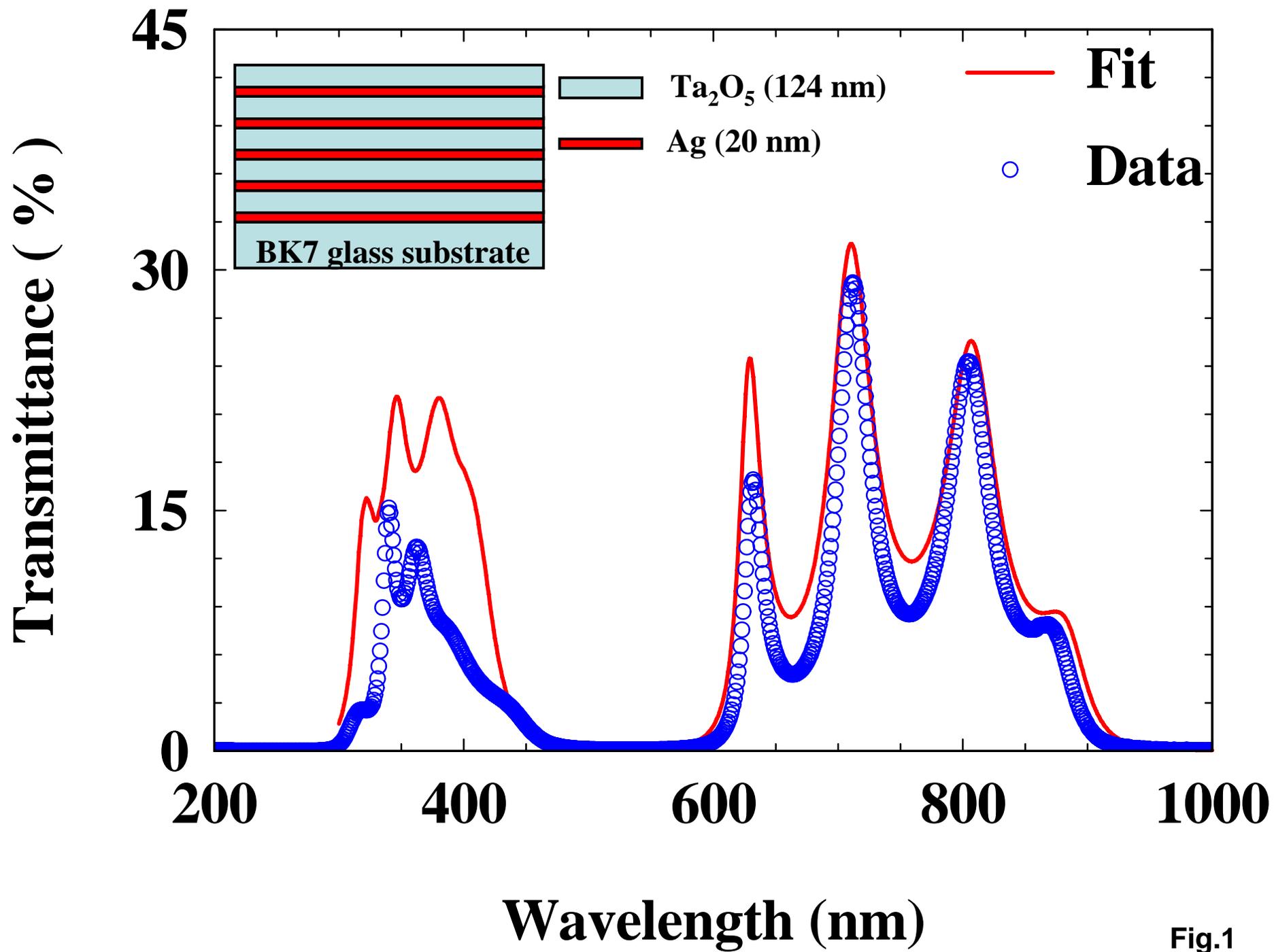

Fig.1

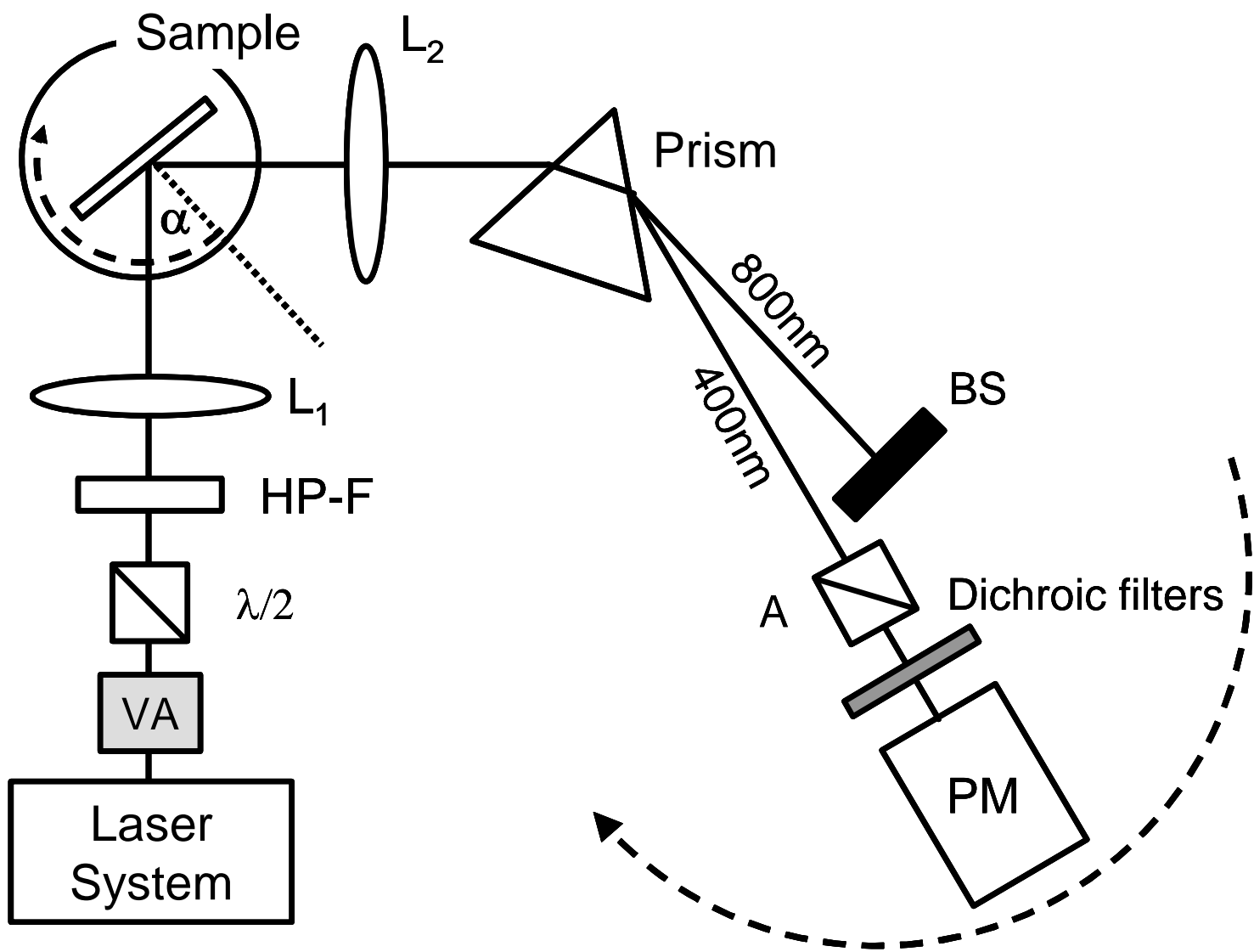

**Fig.2**

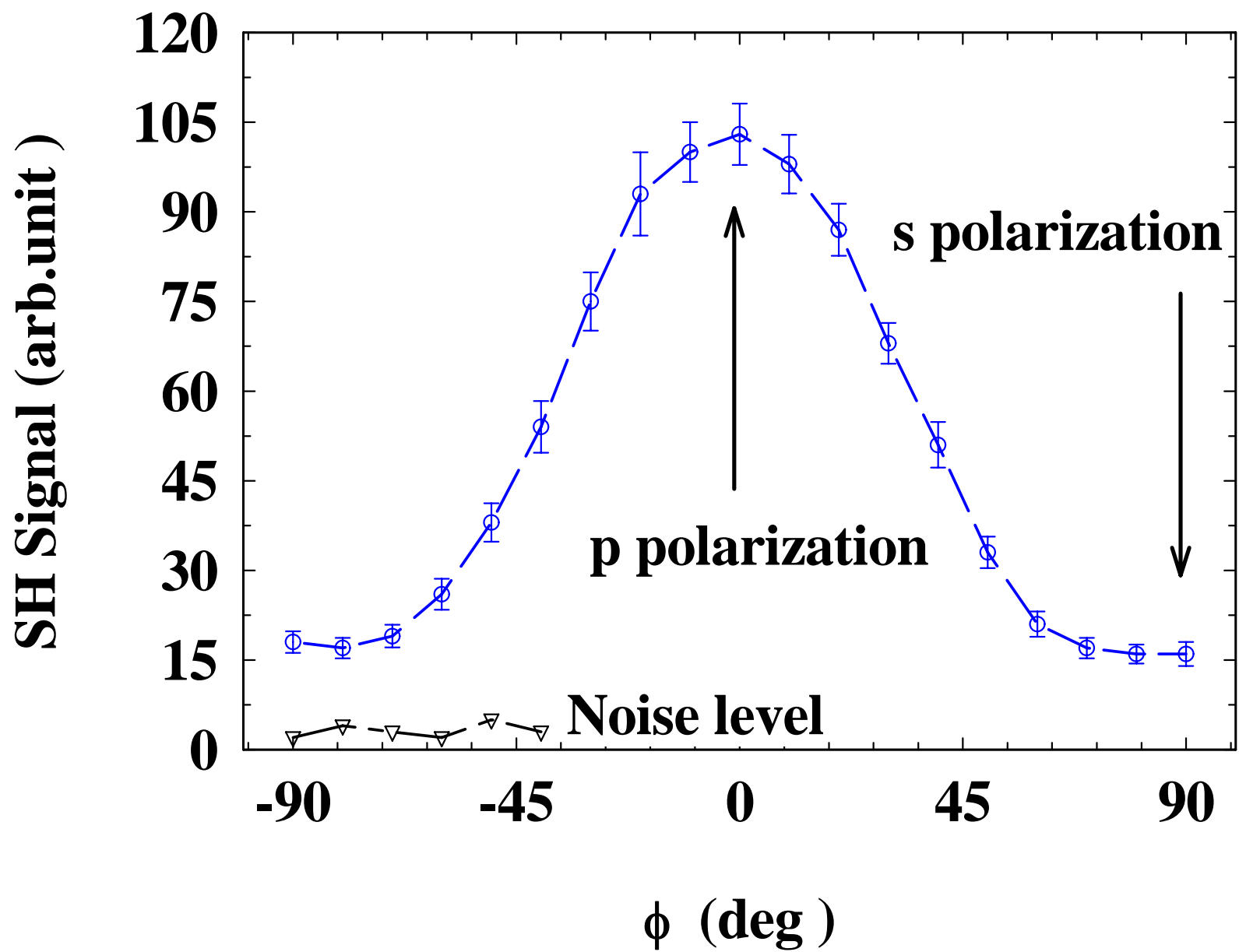

Fig.3

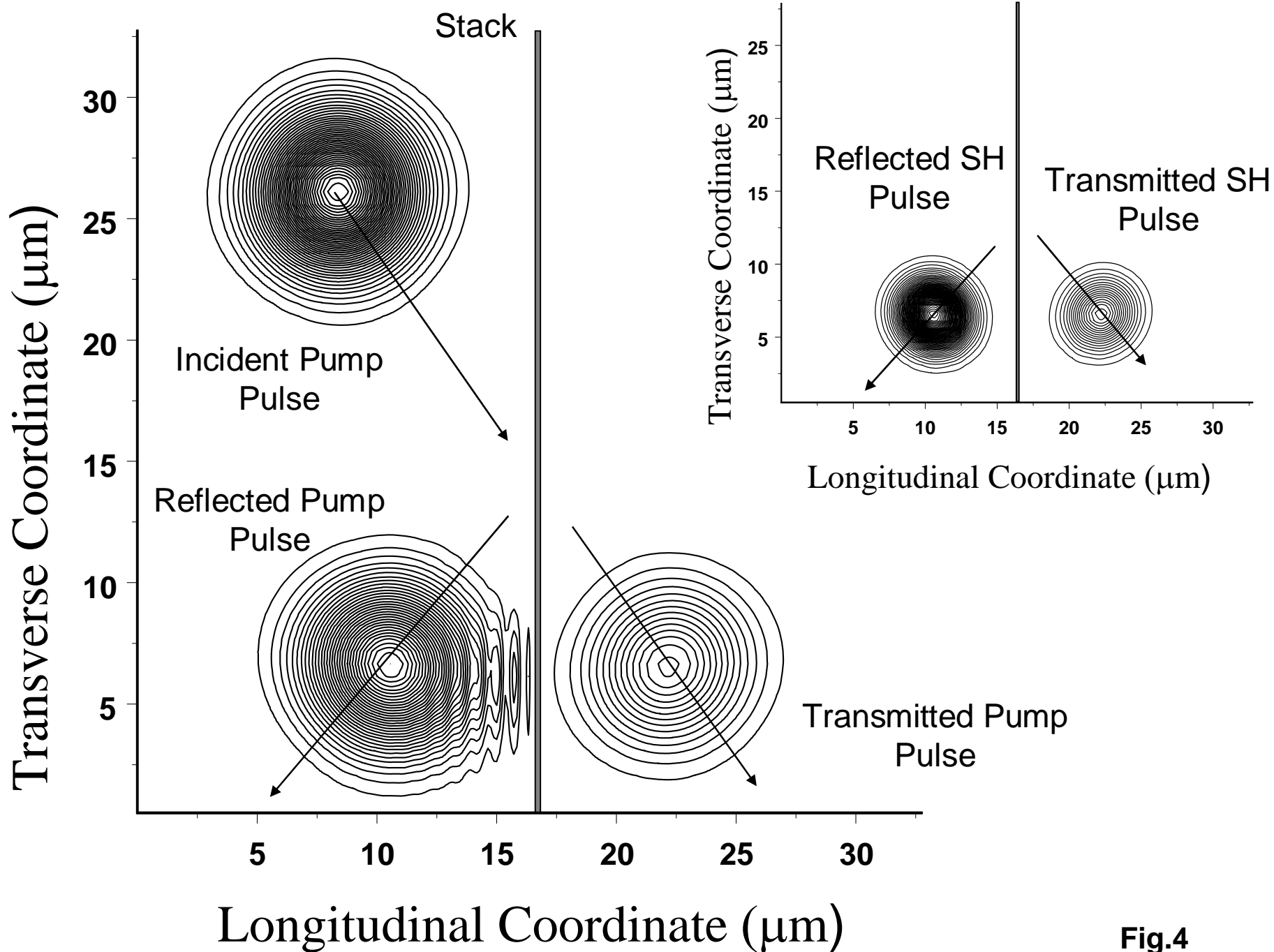

Fig.4

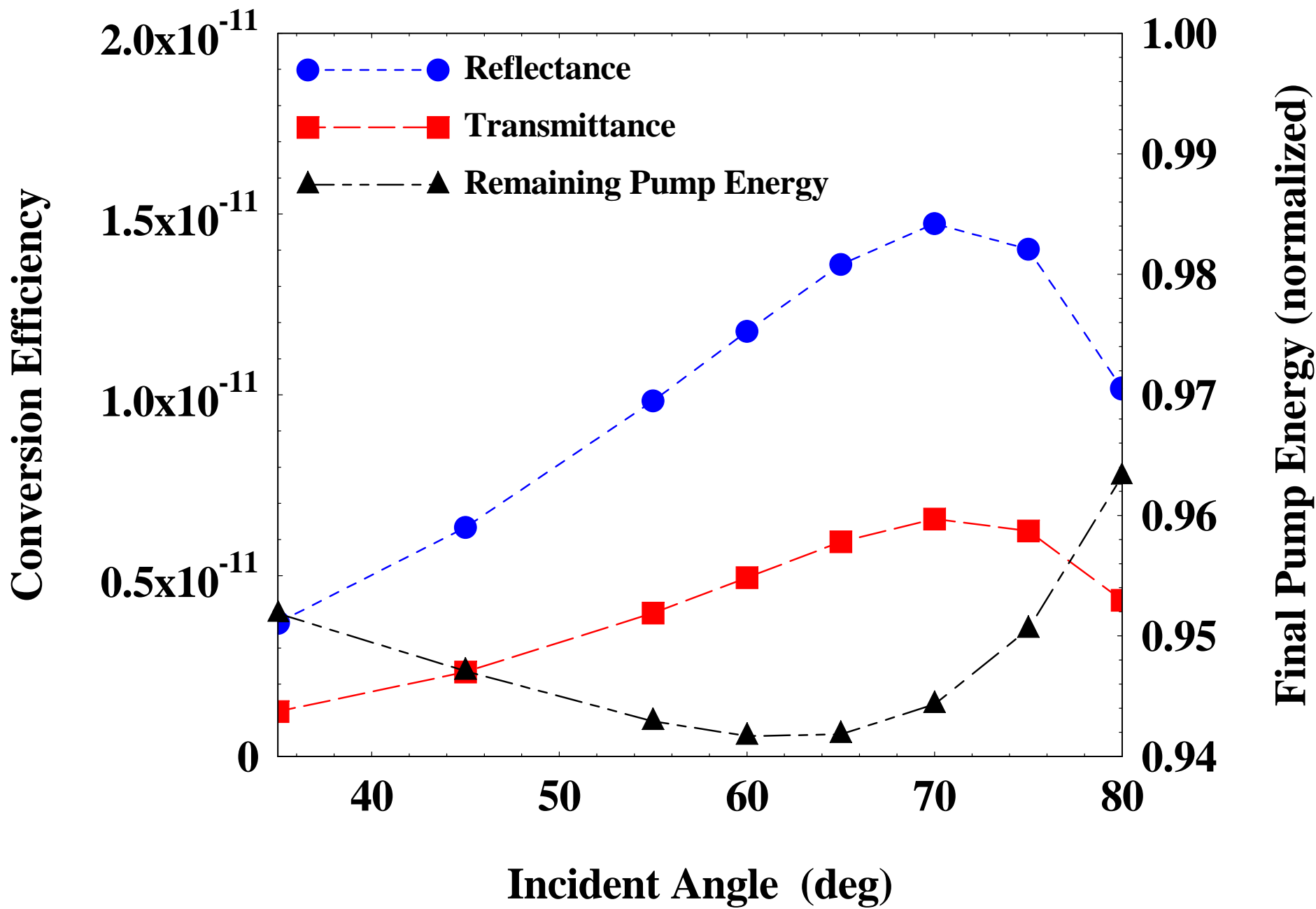

Fig.5

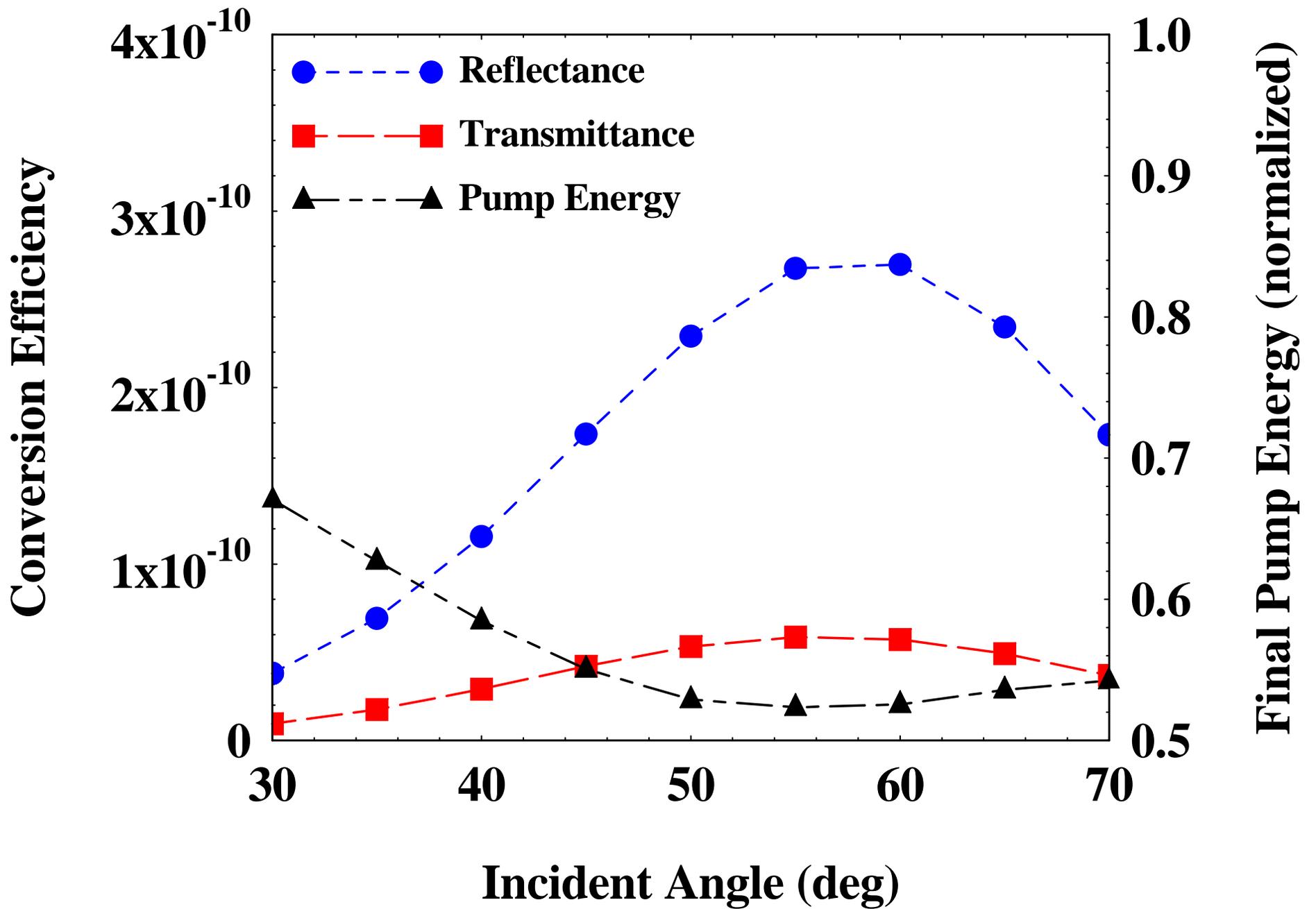

Fig.6

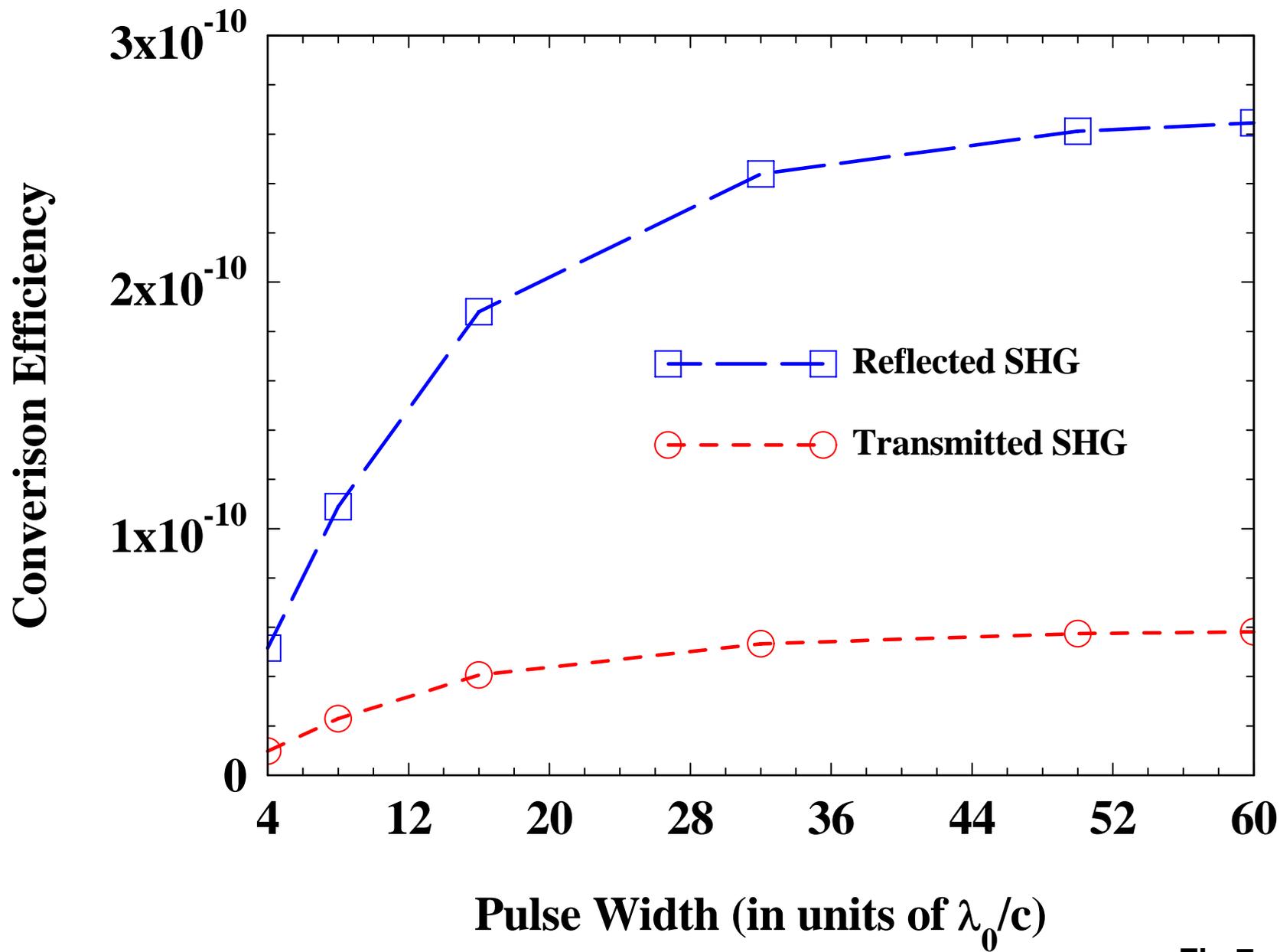

Fig.7

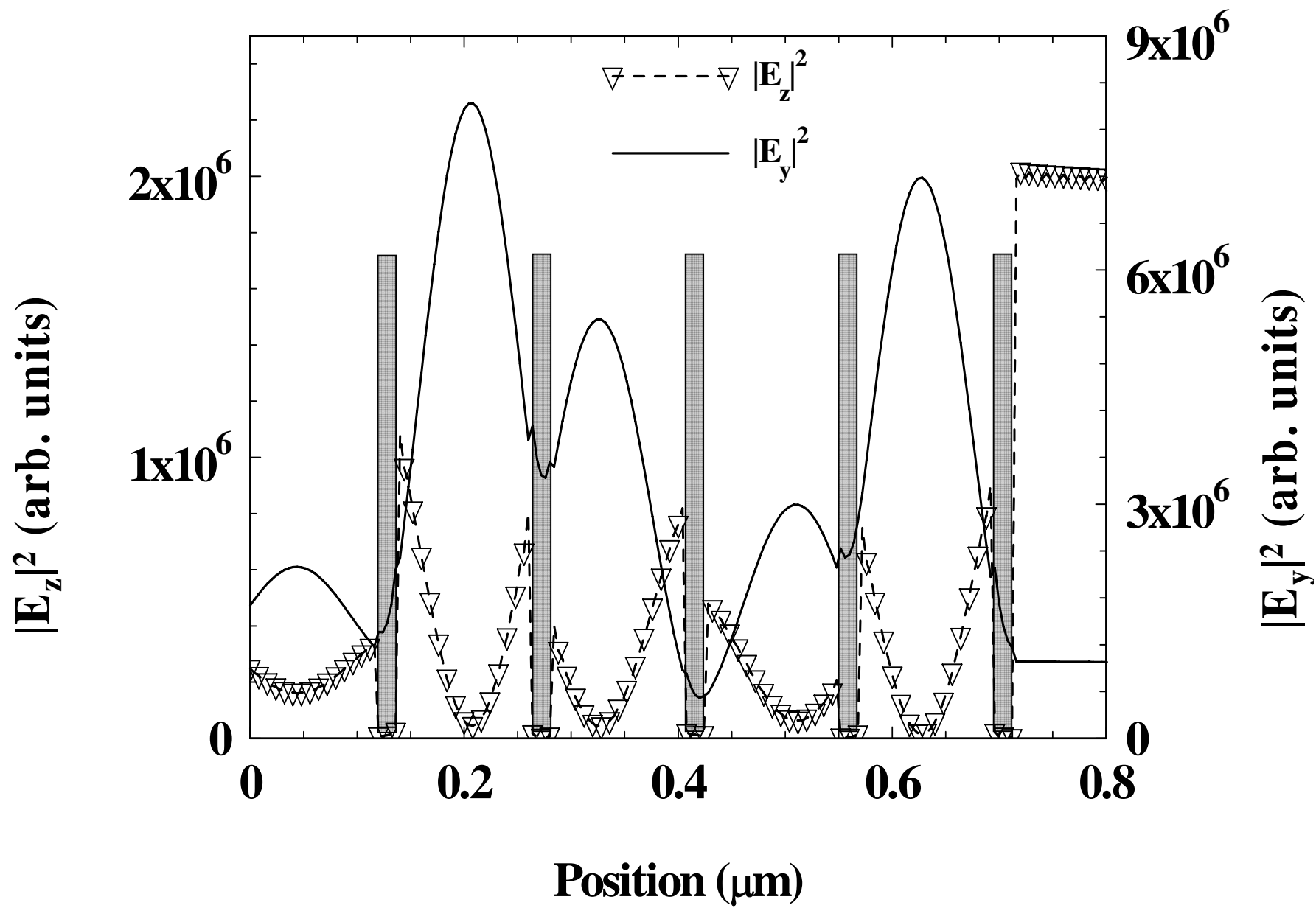

**Fig.8**

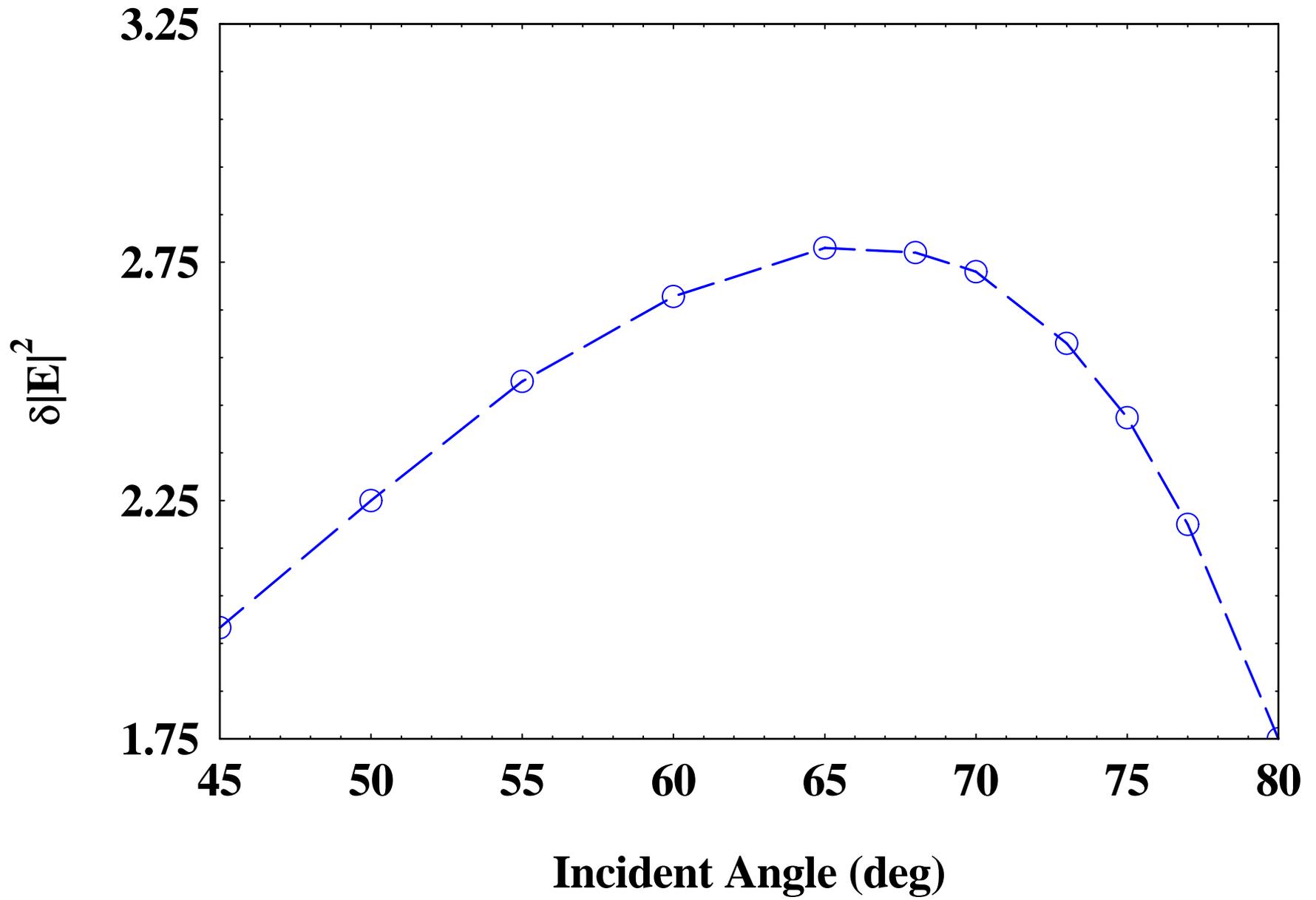

Fig.9

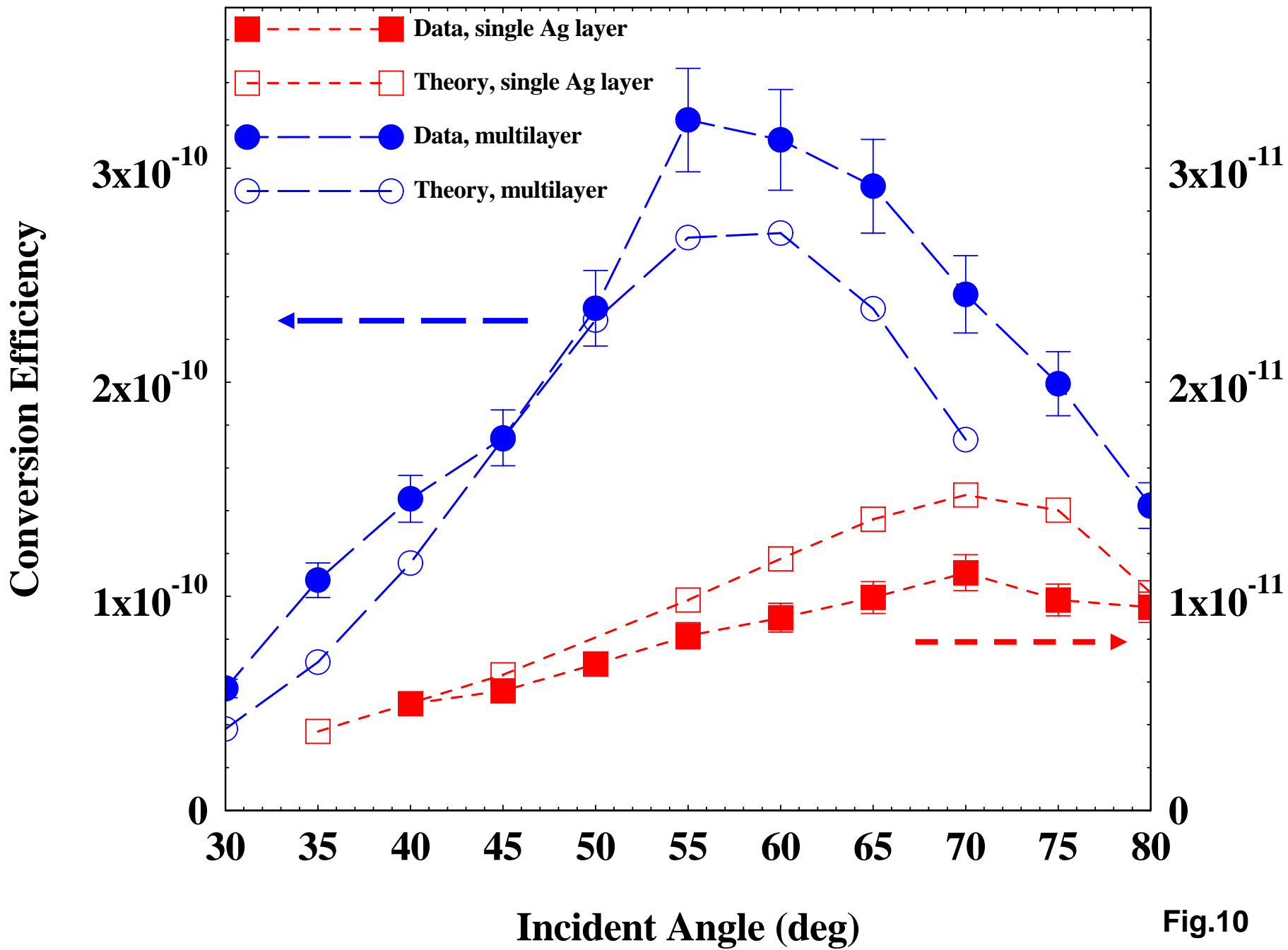

Fig.10